\documentclass[11pt]{article}
\usepackage{moriond,epsfig}

\def\Journal#1#2#3#4{{#1} {\bf #2}, #3 (#4)}

\def\NIMA{{\em Nucl. Instrum. Methods} A}
\def\NPB{{\em Nucl. Phys.} B}
\def\PLB{{\em Phys. Lett.}  B}
\def\PRL{\em Phys. Rev. Lett.}
\def\PRD{{\em Phys. Rev.} D}

\def\numksppcand{5299}
\def\semileplim{$0.87\%$}
\def\wssignif{$4.5$}

\def\ks{$K^{0}_{S}$}
\def\cp{$CP$}\def\pis{$\pi^{+}_{s}$}
\def\dstar{$D^{*+}$}
\def\dz{$D^{0}$}
\def\dzb{$\overline{D^{0}}$}

\def\dzdzb{$D^{0}-\overline{D^{0}}$}

\def\dztokspp{$D^{0}\rightarrow K^{0}_{S}\pi^{+}\pi^{-}$}
\def\dztoksppws{$D^{0}\rightarrow K^{*}(892)^{+}\pi^{-}$}
\def\dztokspprs{$D^{0}\rightarrow K^{*}(892)^{-}\pi^{+}$}
\def\dztoKslnu{$D^{0}\rightarrow K^{*+}e^{-}\overline{\nu}_{e}$}
\def\dsttodpi{$D^{*+}\rightarrow D^{0}\pi_{s}^{+}$}

\def\kstp{$K^{*}(892)^{+}$}
\def\rhoz{$\rho(770)^{0}$}
\def\omegaz{$\omega(792)$}
\def\kstm{$K^{*}(892)^{-}$}
\def\fzero{$f_{0}(980)$}
\def\ftwo{$f_{2}(1270)$}
\def\fzeroprime{$f_{0}(1370)$}
\def\kzero{$K_{0}(1430)^{-}$}
\def\ktwo{$K_{2}(1430)^{-}$}
\def\kstprime{$K^{*}(1680)^{-}$}

\begin{document}
\vspace*{4cm}
\title{RECENT CHARM RESULTS FROM CLEO}

\author{ A. Smith }

\address{Department of Physics, University of Minnesota, 116 Church St. S.E.,\\
Minneapolis, MN, 55455 USA}

\maketitle\abstracts{ We describe two recent results using data
  collected with the CLEO detector at the CESR $e^{+}e^{-}$ collider at
  energies near the $\Upsilon(4S)$.  The first is a Dalitz plot analysis
  of the decay 
  \dztokspp.  We observe a rich structure including the decay
  $D^{0}\rightarrow K^{*+}\pi^{-}$ which may be produced by
  \dzdzb\ mixing or doubly Cabibbo-suppressed decays.  We also search
  for \dzdzb\ mixing in the decay \dztoKslnu.  We observe no events and 
  limit the mixing parameter $R_{\rm mix}$ to be less than \semileplim\ at 
  the 95\% confidence level.
}

\section{Introduction}\label{sec:intro}

\dzdzb\ mixing arises from a mass difference 
or width difference in the mass
eigenstates.  The mixing amplitudes from these contributions are
given by $x \equiv \Delta M/\Gamma_{D}$ and $y \equiv
\Delta\Gamma/2\Gamma_{D}$, where $\Delta M$ and
$\Delta\Gamma$ are the mass and width differences between the mass
eigenstates, respectively, and $\Gamma_{D}$ is the width of the $D$ meson. 
The existence of a massive non-standard model particle could lead to an
observable enhancement of $x$~\cite{bib:largex}.  These measurements are
complementary to 
measurements of mixing in the $K^{0}$ and $B^{0}_{(s)}$ systems in that
only \dzdzb\ mixing would be sensitive to the existence of new down-type
particles. 

These analyses are based on 9~fb$^{-1}$ of $e^{+}e^{-}$ collisions
produced in the CESR collider at energies near the $\Upsilon(4S)$
resonance and recorded using the CLEO~II.V detector~\cite{bib:CLEO}.
We use \dz\ candidates from the decay \dsttodpi\ and use the
charge of the slow pion ($\pi_{s}^{\pm}$) to tag  
whether the initial particle is a \dz\ or \dzb.  The possible final 
states of a \dz\ decay are categorized as ``right signed'' (RS) if they
contain an $s$ quark (like $\overline{K^{0}}$ or $K^{-}$) and ``wrong signed''
(WS) if they contain an $\overline{s}$ quark (like $K^{0}$ or $K^{+}$).  While
not explicitly written, charge conjugate modes are implied here and
throughout this paper.  Cabibbo-favored (CF) decays are
the only contribution to the RS final state.  The WS final state may
reached by $x$ or $y$ mixing or, for hadronic modes, by
doubly Cabibbo-suppressed (DCS) decays.  In hadronic decay channels, the
situation is complicated by a possible strong phase shift
between the CF and DCS channels, which leads to observable variables
$x^{\prime}$ and $y^{\prime}$, related to $x$ and $y$ by
$x^{\prime}=x\cos\delta_{s} + y\sin\delta_{s}$ and
$y^{\prime}=y\cos\delta_{s} - x\sin\delta_{s}$.  Measurement of this
strong phase shift is essential in order to resolve $x$ and
$y$~\cite{Petrov}.  Other WS hadronic decays of \dz\ to $K^{+}\pi^{-}$,
$K^{+}\pi^{-}\pi^{0}$, and $K^{+}\pi^{-}\pi^{+}\pi^{-}$ have been
studied previously by CLEO~\cite{Asner,Smith,Marsh}.

\cp\ violation in charm decays  
is predicted to be very small in the standard model.  If \cp\ violation
were observed in charm decays, it would be strong evidence for new physics.

In addition to \dzdzb\ mixing searches, multi-body $D$ decays can be used study the spectroscopy of
light mesons and glueball candidates using Dalitz plot analyses.
These analyses are only possible with the large statistics $D$ samples
recently available and are complementary to previous studies.

\section{Dalitz Plot Analysis of \dztokspp }\label{sec:D0kspp}

The decay \dztokspp\ may be used to
search for \dzdzb\ mixing and doubly Cabibbo-suppressed decays, 
to study the spectroscopy of light mesons and glueball candidates, and to
understand the effects of final state interactions.  
In this paper we describe our observation of new decay submodes,
including the WS decay \dztoksppws, in a fit of the Dalitz plot.
Because the RS $K^{*}(892)^{-}\pi^{+}$ and WS $K^{*}(892)^{+}\pi^{-}$
intermediate states both decay to a common final state, they may
interfere with each other.  The phase of this interference can be
extracted by fitting the Dalitz plot.  

We reconstruct the \ks\ candidates in the $\pi^{+}\pi^{-}$ channel,
where the pions are required to form a common vertex with a confidence
level greater than $10^{-6}$.  We utilize the
precision tracking of the CLEO~II.V silicon vertex detector
(SVX)~\cite{SVX} to refit the 
\ks, $\pi^{+}$, and $\pi^{-}$ tracks to a common vertex and require the
confidence level of the fit to exceed $10^{-4}$.  The \pis\ track is
refit with the constraint that it originate from the intersection of the
\dz\ candidate trajectory and the CESR luminous region.

In addition to the \dz\ candidate mass, we also reconstruct the energy
release in the \dstar\ decay, $Q\equiv M^{*}-M-m_{\pi}$, where $M^{*}$
is the mass of the \dstar\ candidate, $M$ is the mass of the \dz\
candidate, and $m_{\pi}$ is the mass of the charged pion.  The
distribution of $Q$ has a core width of $220\pm 4$~keV.
The momentum of the \dstar candidate is required to be greater the 2.0
GeV/$c$.   
Our data selection results in \numksppcand\ candidate events within the
signal region of three standard deviations about the central $Q$, $M$,
and $m(\pi^{+}\pi^{-})$ values.  The fraction of background in the
signal region is found to be quite small ($2.1\pm 1.5\%$) using a fit
to $M$.  

The Dalitz plot variables used in the fit are $M^{2}(\pi^{+}\pi^{-})$
and  $M_{RS}$, where $M_{RS} = M(K_{0}^{S}\pi^{-})$ for $D^{0}$ decays
and $M_{RS} = M(K_{0}^{S}\pi^{+})$ for $\overline{D^{0}}$ decays.  
The decay to a three-body $K^{0}_{S}\pi^{+}\pi^{-}$ final state is
modeled as a pseudo-two-body decay in which one daughter is
a resonance, followed by a two-body decay of the resonance.
The Dalitz plot is fitted to the square of 
the sum of resonant and non-resonant amplitudes multiplied by complex
coefficients.  Each resonant component 
is modeled by a relativistic Breit-Wigner amplitude multiplied by
Blatt-Weisskopf form factors~\cite{Blatt} for the \dz\ and intermediate
particle decay vertices and a factor to give the correct $J$-dependent
angular distribution. 

The efficiency function is determined using a sample of non-resonant Monte
Carlo, and is found to be nearly uniform across the Dalitz plot.  The
Dalitz plot of the small background is determined 
using sideband regions five to ten standard deviations away from the
signal in $Q$ and $M$.  These distributions are fitted to a
two-dimensional polynomial and uncertainties are
determined by varying the parameters by one standard deviation.

\begin{figure}
  \centerline{
    \epsfysize=2.25in
    \epsffile{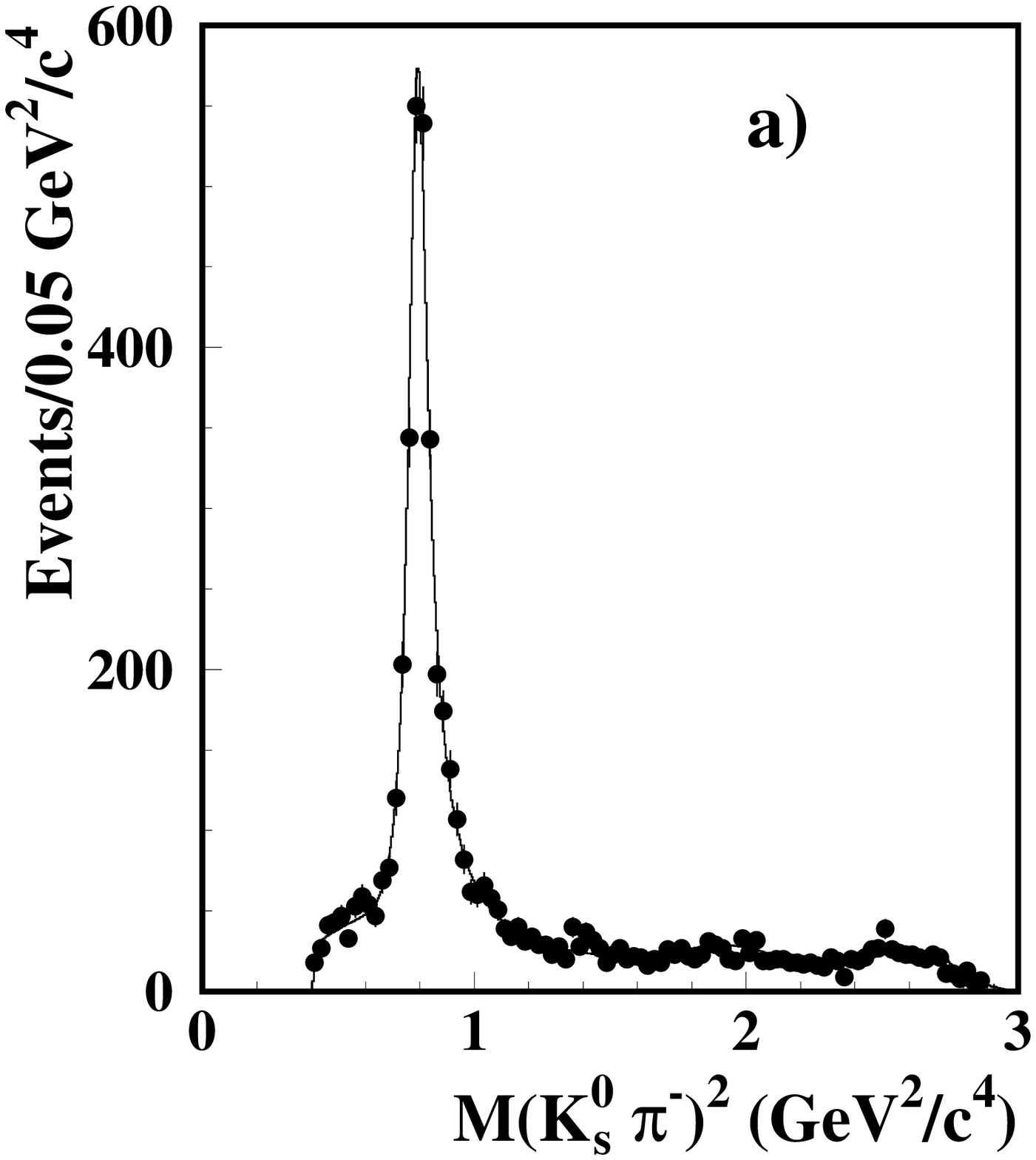}
    \epsfysize=2.25in
    \epsffile{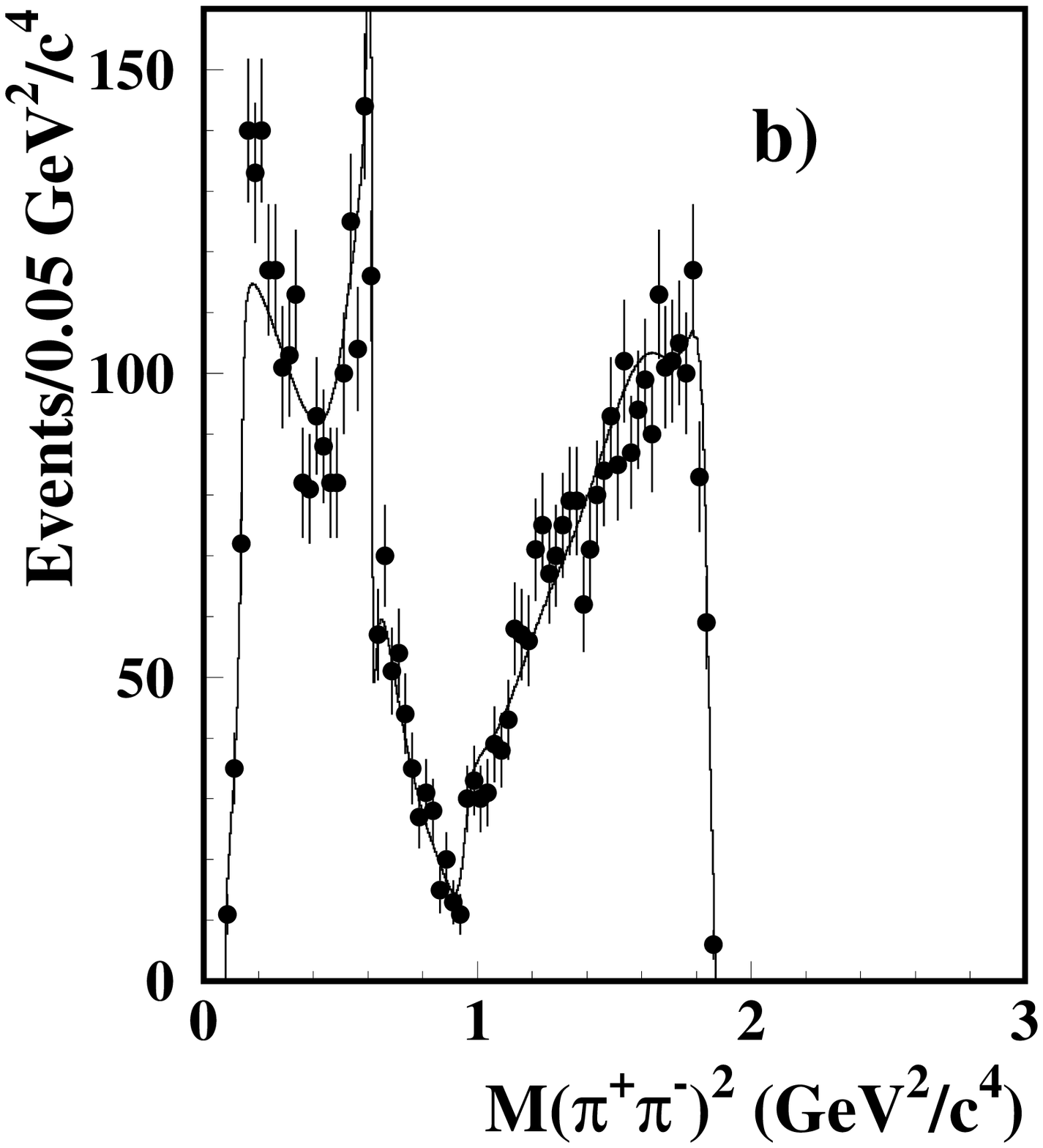}
    \epsfysize=2.25in
    \epsffile{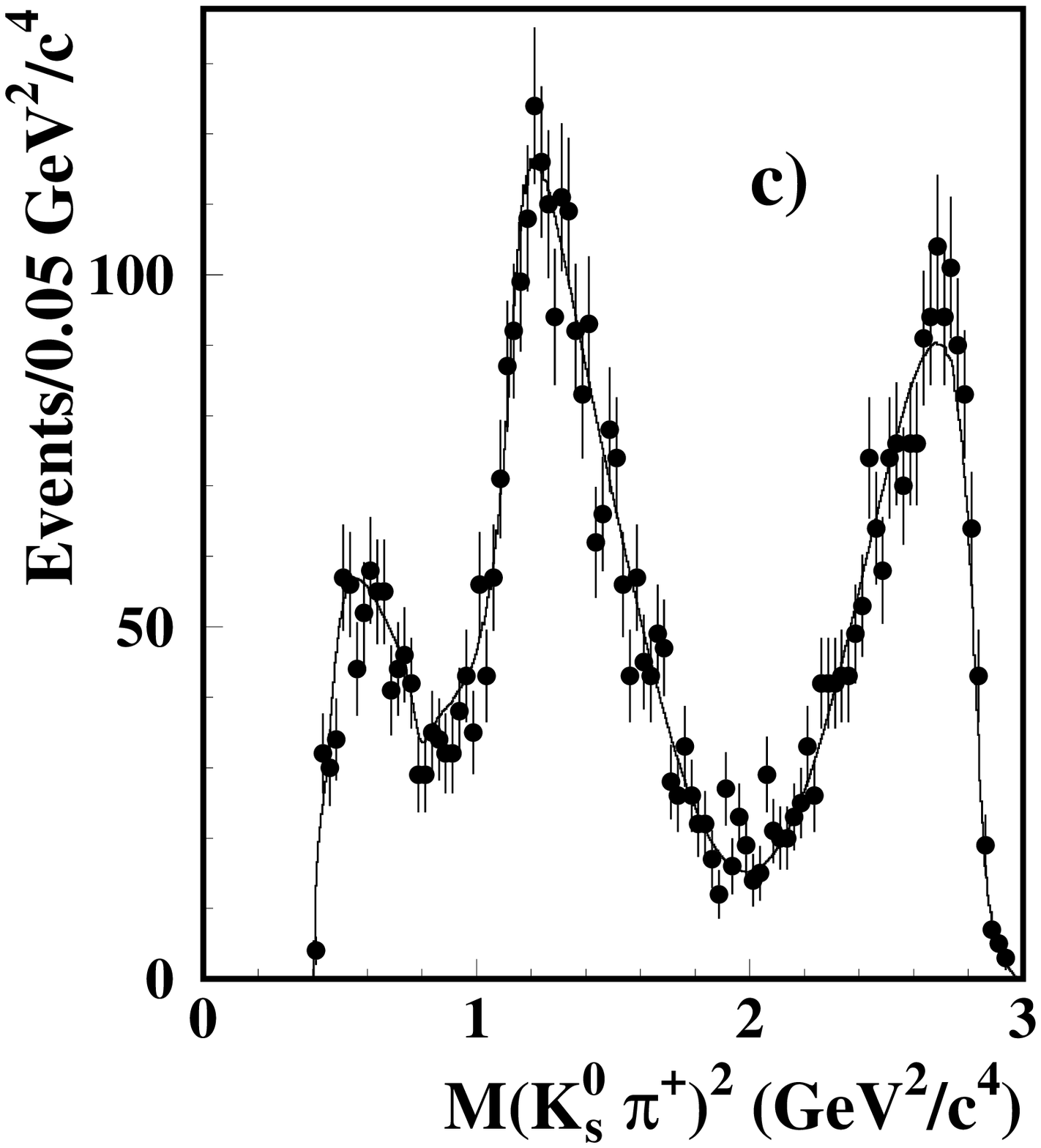}
  }
  \caption{Projections of the Dalitz plot of \dztokspp\ in a) $M^{2}(K^{0}_{S}\pi^{-})$, b)
  $M^{2}(\pi^{+}\pi^{-})$, and c) $M^{2}(K^{0}_{S}\pi^{+})$ showing the 
  data (points) and fit function (line).\label{fig:dalfit}}
\end{figure}

The Dalitz plot is fitted using a maximum likelihood technique.  We
consider non-resonant and 18 possible resonant contributions: \kstm,
$K_{0}(1410)^{-}$, \kzero, \ktwo, \kstprime, $K_{3}(1780)^{-}$, \omegaz,
\rhoz, $\rho(1450)^{0}$, $\rho(1700)^{0}$, $\sigma(500)$, \fzero, \ftwo,
\fzeroprime, $f_{0}(1500)$, $f_{0}(1710)$, \kstp, $K_{0}(1430)^{+}$.  
The Dalitz plot and projections showing the fit results plotted against
the data are shown in Fig.~\ref{fig:dalfit}.
The results of the nominal fit are shown in Tab.~\ref{tab:dalfit}.  
An amplitude for the WS decay mode \dztoksppws\ is observed with a
statistical significance of \wssignif\ standard deviations.  This is the
first observation of this WS decay.

Systematic uncertainties on amplitudes and phases arise from
uncertainties in the choice of resonances used in the fit, modeling
of the decays, and experimental limitations.  In particular, there are
uncertainties associated with our poor understanding of $\pi\pi$ scalar
resonances and threshold effects.  The fit strongly prefers a
feature consistent with the $\sigma(500)$, however, our 
Breit-Wigner parameterization of this contribution is inadequate for the
purpose of establishing its existence.  A partial wave analysis would be
required.  We try several fits with different contributions
included or excluded in order to estimate the systematic uncertainty on
the WS contribution.

We observe the WS decay mode \dztoksppws\ and measure its branching
fraction and strong phase shift relative to the RS decay \dztokspprs\ to
be $R_{WS}=(0.6 \pm 0.3 \pm 0.2)\%$ and $(-3 \pm 11 \pm 8)^{\circ}$,
respectively.  \dz\ and \dzb\ decays are also analyzed separately
and no $CP$ violating effects are observed at the few percent level.

\renewcommand{\arraystretch}{1.2}
\begin{table}[t]
\caption{Results of the Dalitz plot fit of the decay \dztokspp .  The
  first uncertainty is statistical, the second is the experimental
  systematic uncertainty, the third is due the choice of
  the resonance model and which resonances are included in the fit.
\label{tab:dalfit}}
\vspace{0.4cm}
\begin{center}
\begin{tabular}{|c|c|c|c|}
\hline
Submode & Amplitude & Phase ($^{\circ}$)& Fit Fraction \\
\hline
\kstp & $0.131 \pm 0.029\ ^{+0.050}_{-0.016}\ ^{+0.054}_{-0.030}$ & $152
\pm 11\ ^{+5}_{-8}\ ^{+4}_{-11}$ & $0.0041 \pm 0.0020$ \\
\rhoz & $1.0$ (fixed) & $0$ (fixed) & $0.227 \pm 0.011$ \\
\omegaz & $0.055 \pm 0.006 \pm 0.01\ ^{+0.005}_{-0.006}$ & $117 \pm 7 \pm
4\ ^{+3}_{-8}$ & $0.013 \pm 0.003$ \\
\kstm & $1.68\pm 0.05 \pm 0.04\ ^{+0.05}_{-0.07}$ & $149 \pm 2\
^{+3}_{-2}\ ^{+3}_{-5}$ & $0.672 \pm 0.014$\\
\fzero & $0.36 \pm 0.02\ ^{+0.06}_{-0.03}\ ^{+0.06}_{-0.02}$ & $185 \pm
5\ ^{+1}_{-3}\ ^{+19}_{-9}$ & $0.040 \pm 0.005$ \\
\ftwo & $0.8 \pm 0.2\ ^{+0.2}_{-0.3}\ ^{+0.8}_{-0.2}$ & $368 \pm 21 \pm 13
\pm 27$ & $0.006 \pm 0.002$ \\
\fzeroprime & $1.56 \pm 0.14\ ^{+0.24}_{-0.010}\ ^{+0.7}_{-0.5}$ & $74 \pm
6\ ^{+2}_{-17}\ ^{+21}_{-11}$ & $0.07 \pm 0.01$ \\
\kzero & $2.2 \pm 0.1\ ^{+0.1}_{-0.2} \pm 0.4$ & $11 \pm 5\ ^{+6}_{-5}\ 
^{+10}_{-22}$ & $0.080 \pm 0.009$\\
\ktwo & $0.9 \pm 0.1 ^{+0.3}_{-0.0} \pm 0.4$ & $-47 \pm 8 \pm 6
^{+19}_{-11}$ & $0.015 \pm 0.002$ \\
\kstprime & $5.6 \pm 0.8 \pm 0.9 ^{+2.8}_{-4.2}$ & $173 \pm 8 \pm 12
^{+18}_{-20}$ & $0.023 \pm 0.006$ \\
Non-resonant & $1.9 \pm 0.6\ ^{+0.6}_{-0.4}\ ^{+1.4}_{-0.9}$ & $-30 \pm
11\ ^{+16}_{-7}\ ^{+16}_{-7}$ & $0.027 \pm 0.016$\\
\hline
\end{tabular}
\end{center}
\end{table}
\renewcommand{\arraystretch}{1.0}

\section{Limit on \dzdzb\ Mixing Using \dztoKslnu }\label{sec:D0kslnu}

The decay mode \dztoKslnu\ is sensitive to $R_{mix}\equiv
\frac{1}{2}(x^{2}+y^{2})$, but not to $x$ directly.  While there are
experimental difficulties related to the undetected neutrino and soft
spectra of the daughter tracks, only mixing contributes to the WS final
state.

The $K^{*+}$ is reconstructed in the $K^{0}_{S}\pi^{+}$ mode, where the
$\pi^{+}\pi^{-}$ final state is used to reconstruct the $K^{0}_{S}$.
The neutrino is ``pseudo-reconstructed'', taking advantage of the
hermeticity of the CLEO detector in order to improve the
resolution on $Q$.  The direction of the neutrino is taken from a
weighted average of the event thrust axis, the slow pion direction, and
the direction of all \dz\ daughters other than the neutrino.  The
optimal weights of these measurements is based on Monte Carlo
simulations.  The remaining quadratic ambiguity on the momentum is
resolved by considering the choice which gives the most probable
combination of \dz\ momentum and electron decay angle in the $W$ rest
frame.  

A maximum likelihood fit to the candidate $Q$ and proper time is used
to determine the signal contribution.  The proper time for signal is
expected to have a $t^{2}e^{-t}$ dependence, where $t$ is the proper
time normalized to the $D^{0}$ lifetime.  The fit to the RS data yields
$638 \pm 51$ events.  The WS fit yields $0.00\pm 1.99$ events, leading
to an upper limit on $R_{mix}$ of $<$\semileplim\ at $95\%$ confidence
level.

\section{Outlook}\label{sec:summary}

Several searches for \dzdzb\ mixing and studies of light meson
spectroscopy using Dalitz analyses of $D$ decays are underway at CLEO.
We expect to have public results of a {\em time-dependent} Dalitz analysis of
\dztokspp\ in the near future .  Since $\delta_{s}$ is measured, $x$ and $y$ are measured
directly.  Furthermore, this
is the only analysis with sensitivity to the sign of $x$.   
The \dztoKslnu\ result will be combined with results of an analysis of
$D^{0}\rightarrow K^{+}\ell^{-}\overline{\nu}_{\ell}$ which is underway.  

\section*{Acknowledgments}

We gratefully acknowledge the effort of the CESR staff in providing us with
excellent luminosity and running conditions.  This work was supported by 
the National Science Foundation, the U.S. Department of Energy,
the Research Corporation, and the Texas Advanced Research Program.

\end{document}